\documentclass[reprint,showpacs,amsmath,amssymb,aps,usenatbib]{revtex4-1}
\usepackage{graphicx}
\usepackage{dcolumn}
\usepackage{bm}
\usepackage{subfigure}
\usepackage{longtable}
\usepackage{setspace}
\usepackage[bookmarksnumbered,bookmarksopen,colorlinks,citecolor=blue,linkcolor=blue]{hyperref}
\usepackage{natbib}
\usepackage{amsmath}
\usepackage{rotating}
\usepackage{colortbl}
\usepackage{color}
\usepackage{epstopdf}
\definecolor{mygray}{gray}{0.8}

\begin{document}
\title{Shell quenching in nuclear charge radii based on Monte Carlo dropout Bayesian neural network}

\author{Zhen-Yan Xian}
\affiliation{School of Physics, Ningxia University, Yinchuan 750021, China}

\author{Yan Ya}
\affiliation{School of Physics, Ningxia University, Yinchuan 750021, China}

\author{Rong An}
\email[Corresponding author: ]{rongan@nxu.edu.cn} 
\affiliation{School of Physics, Ningxia University, Yinchuan 750021, China}
\affiliation{Guangxi Key Laboratory of Nuclear Physics and Technology, Guangxi Normal University, Guilin, 541004, China}
\affiliation{Key Laboratory of Beam Technology of Ministry of Education, College of Nuclear Science and Technology, Beijing Normal University, Beijing 100875, China}

\begin{abstract}
 Charge radii can be generally used to encode information about various fine structures of finite nuclei.
  In this work, a constructed Bayesian neural network based on the Monte Carlo dropout approach is proposed to accurately describe the charge radii of nuclei with proton number $Z\geq20$ and mass number $A\geq40$. More motivated underlying mechanisms are incorporated into this combined model in addition to the basic building blocks with the specific number of protons and neutrons, which naturally contain the pairing effect, the isospin effect, the shell closure effect associated with the Casten factor $P$, the valence neutrons, the valence protons, the quadrupole deformation $\beta_{20}$, the high order hexadecapole deformation $\beta_{40}$, and the local shape staggering effect of $^{181,183,185}$Hg. To avoid the distorted cases of the traditional Casten factor at the fully filled shells, the modified Casten factor $P^{*}$ is introduced into the input structure parameter sets. The standard root-mean-square deviation is reduced to $0.0084$ fm for the training data set and $0.0124$ fm for the validation data set with the modified Casten factor $P^{*}$. Meanwhile, the shell closure effect of nuclear charge radii can be reproduced remarkably well. We have successfully demonstrated the ability of this constructed model to significantly increase the accuracy in predicting the nuclear charge radii.
\end{abstract}

\maketitle

\section{Introduction}
\label{sec1}
Finite nuclei, characterized as strongly correlated many-body systems, provide a natural probe for investigating fundamental interactions and forces.
As directly measured quantities in atomic nuclei, charge radii can be generally used to reflect the range of the proton density distributions and encode information about various fine structure phenomena, such as the emergence of shell closure effects~\cite{Angeli_2015,PhysRevC.100.044310,GarciaRuiz:2019cog,PhysRevC.104.064313,PhysRevLett.129.142502} and the odd-even staggering (OES) behavior~\cite{GarciaRuiz:2016ohj,Marsh:2018wxs,Miller2019,PhysRevLett.127.192501}.
Furthermore, the available description of the proton density distributions can also be used to pin down the components of the isospin interactions in the equation of state (EoS) of asymmetric nuclear matter~\cite{PhysRevC.88.011301,PhysRevLett.119.122502,PhysRevResearch.2.022035,PhysRevLett.127.182503,nuclscitech34.119,PhysRevLett.132.162502}.
This means that more data on nuclear charge radii should be urgently required with highly accurate precision.
Remarkable achievements in detecting charge radii have been achieved over the past decades.
Especially for nuclei with extreme neutron-to-proton ratios, the laser spectroscopy method provides a credible access to measure the nuclear size~\cite{CAMPBELL2016127,YANG2023104005}.


The undertaken efforts have been devoted to featuring the normal and abnormal variations in nuclear charge radii along a long isotopic family, which includes the empirical formulas~\cite{Zhang2002EPJA285,PhysRevC.87.024310,Piekarewicz2010,PhysRevC.94.064315}, relativistic mean field theory~\cite{Geng:2003pk,PhysRevC.102.024307,PhysRevC.109.064302}, non-relativistic Skyrme Hartree-Fock-Bogoliubov (HFB) approach~\cite{PhysRevLett.102.242501,PhysRevC.82.035804}, and the Fayans energy density functional~\cite{PhysRevC.95.064328,PhysRevLett.122.192502}, etc.
Recently, machine learning methods have been made great applications in the course of nuclear
physics~\cite{PhysRevC.98.034318,Bedaque2021,Paquet_2024}.
Particularly, a well-known approach based on Bayesian neural network (BNN) has been widely and successfully employed to describe various nuclear quantities from finite nuclei to infinite nuclear matter, which includes the nuclear mass~\cite{PhysRevC.93.014311,PhysRevC.96.044308,NIU201848,PhysRevC.100.054311,Kejzlar_2020,PhysRevC.101.051301,
PhysRevC.104.014315,PhysRevC.106.014305,PhysRevC.106.L021303,ZHANG2024122820},
$\alpha$-decay properties~\cite{Rodríguez_2019,PhysRevC.107.014310,PhysRevC.108.014326,PhysRevC.110.024319} and $\alpha$-clustering structure~\cite{PhysRevC.104.044902},
$\beta$-decay properties~\cite{PhysRevC.99.064307,PhysRevC.106.024306,PhysRevC.107.034316} or
double-$\ensuremath{\beta}$-decay models~\cite{PhysRevD.96.053001,PhysRevD.104.055040,PhysRevC.94.024603}, $\gamma$-ray emission~\cite{PhysRevC.105.045801}, nuclear fission~\cite{Schunck_2020}, low-lying excited phenomena~\cite{PhysRevC.104.034317,WANG2022137154}, landscape of nuclide chart~\cite{PhysRevC.101.044307,PhysRevLett.122.062502}, neutron skin thickness~\cite{PhysRevC.107.045802,PhysRevC.108.024317}, the yields of production cross sections~\cite{PhysRevC.107.064909,PhysRevC.108.044606,YingHuaDang}, and nuclear charge radii~\cite{Utama_2016,PhysRevC.101.014304,PhysRevC.102.054323,PhysRevC.105.014308,DONG2023137726}, etc. In addition, Bayesian inference can help to gain deeper insights into the understanding of fundamental interactions~\cite{PhysRevC.105.044305,Wesolowski_2016,PhysRevC.97.025805,YANG2020135540,PhysRevC.103.025807,XU2020135820,PhysRevC.103.065804,
PhysRevC.105.014004}, significantly the bulk properties of nuclear matter~\cite{PhysRevC.96.065805,PhysRevLett.125.202702,
PhysRevC.103.064323,PhysRevD.103.063036,Xie_2021,PhysRevC.105.015806,PhysRevC.106.044305,PhysRevD.106.043024,
panglonggang,PhysRevC.107.045803,PhysRevC.108.034908,QIU2024138435,RADUTA2024138696,PhysRevC.107.055803,PhysRevC.110.035805,
PhysRevD.109.123038}.
It should be mentioned that Bayesian neural networks can cover the quantitative uncertainties associated with these various physical systems~\cite{Higdon2015,PhysRevC.96.024003,PhysRevC.98.024305,PhysRevLett.119.252501,HU2019134982,
PhysRevC.101.055803,PhysRevC.102.024616,PhysRevC.102.054315,Lovell_2020,PhysRevC.104.054324,PhysRevC.106.024607,
PhysRevC.106.054001,PhysRevC.108.024321,PhysRevC.109.054301,XinyuWang2024}.
This encourages us to perform the proceeding research in exploring the systematic evolution of nuclear charge radii through Bayesian neural network-based approach.

An accurate description of nuclear charge radii can be applied to refine the local variations in nuclear size.
As demonstrated in Ref.~\cite{Utama_2016}, BNN enhances the prediction ability of nuclear charge radii by training the residuals between the theoretical predictions and the experimental data. This leads to the deviation of the root-mean-square (rms) around $0.02$ fm.
The improved BNN that considers the pairing and shell closure effects gives the rms deviation $0.015$ fm in calibrating the charge radii data of nuclei with mass number $A\geq40$ and proton number $Z\geq20$~\cite{PhysRevC.105.014308}.
As is well known, the larger odd-even staggering in the charge radii can be profoundly observed in the neutron-deficient isotopes around the proton number $Z=82$ isotopic chains~\cite{Marsh:2018wxs,PhysRevLett.127.192501,PhysRevC.95.044324,ANSELMENT1986471,PhysRevC.104.024328,PhysRevLett.126.032502}.
This is attributed to the underlying mechanism of the abrupt shape-phase transition.
Considering the abnormal shape-staggering effect, the standard rms deviation falls into $0.014$ fm for both the trained and validated charge radii data through the refined Bayesian neural network~\cite{DONG2023137726}.
In addition, different kinds of machine learning methods, such as the naive Bayesian probability classifier~\cite{PhysRevC.101.014304} and artificial neural networks~\cite{PhysRevC.102.054323,Akkoyun_2013,PhysRevC.108.034315}, can be used to explore the systematic evolution of nuclear charge radii as well.

As mentioned above, Bayesian neural networks can capture the considered error quantification by comparing the experimental data and theoretical simulations.
As mentioned in Ref.~\cite{Zhang10662945}, the aleatoric and epistemic uncertainties should be explicitly considered in a given Bayesian neural network.
This means that more underlying physical mechanisms should be captured as much as possible in the calibrated protocol.
To further reduce the quantified uncertainty in validating the nuclear charge radii, both quadrupole deformation $\beta_{20}$ and higher moment deformation $\beta_{40}$ are incorporated into the simulated codes in this work.
The Casten factor $P$, which is characterized as the correlation between the valence neutrons $\nu_{n}$ and the valence protons $\nu_{p}$, is associated with the shell closure effect~\cite{PhysRevLett.58.658,Angeli_1991,Casten_1996}.
In our calculations, the Casten factor $P$, valence neutrons $\nu_{n}$, and valence protons $\nu_{p}$ are simultaneously considered as input parameter sets.

Actually, the Casten factor $P$ plays an indispensable role in the calibrated protocol~\cite{PhysRevC.105.014308,DONG2023137726,Sheng:2015poa}.
For the calculation of the Casten factor $P$, the reference neutron and proton magic numbers are taken as $Z=2$, $6$, $14$, $28$, $50$, $82$, $(114)$ and $N=2$, $8$, $14$, $28$, $50$, $82$, $126$, $(184)$~\cite{PhysRevC.88.011301,Angeli_1991,Sheng:2015poa}.
However, the traditional Casten factor, which is defined as $P=\nu_{n}\nu_{p}/(\nu_{n}+\nu_{p})$, is distorted due to the fully filled proton magic shells or the neutron magic numbers.
In order to avoid the aforementioned problem, the shell closure effect has been taken into account by using the modified Casten factor $P^{*}$.
In this work, the Monte Carlo dropout Bayesian neural network approach (MC-dropout BNN) is built appropriately to describe the charge radii of nuclei with proton number $Z\geq20$ and mass number $A\geq40$.
The input structures include the proton number ($Z$), the mass number ($A$), the pairing effect ($\delta$), the isospin effect ($I^{2}$), the shell closure effect associated with the modified Casten factor $P^{*}$, the valence neutrons $\nu_{n}$, the valence protons $\nu_{p}$, the quadrupole deformation $\beta_{20}$, the high order hexadecapole deformation $\beta_{40}$ and the local `abnormal' shape staggering effect of $^{181,183,185}$Hg ($LI$).
The results obtained by incorporating the traditional Casten factor $P$ into the input structures are also shown for comparison.

The structure of the paper is the following.
In Sec.~\ref{sec2}, the theoretical framework about the MC-dropout BNN approach is briefly presented, which includes the engineered features and the training and validation data sets in the nuclear charge radii.
In Sec.~\ref{sec3}, the numerical results and discussion are provided.
Finally, a summary is given in Sec.~\ref{sec4}.

\section{ Theoretical formalism}
\label{sec2}

The Bayesian neural network (BNN), which is characterized as the combination of an artificial neural network (ANN) and Bayesian statistical theory, provides a potential approach to comprehend the physical quantities with the associated uncertainty.
However, as demonstrated in Ref.~\cite{JMLRv15srivastava14a}, the over-fitting puzzle can be encountered in the traditional neural network.
Then the proposed dropout approach can be used to avoid over-fitting problem of the neural network parameters through the regularization technique.
Meanwhile, the constructed Monte Carlo dropout method (MC-dropout) is being equivalent to variational inference Bayesian learning method in evaluating a neural network~\cite{Gal2016}.
Moreover, this probabilistic model is more time saving compared to the generally used energy density functionals (EDFs). In contrast to the traditional BNN, the Monte Carlo dropout Bayesian neural network (MC-dropout BNN) model can save more computing resources as well.
Therefore, the MC-dropout BNN model has been built based on these motivations.
This approach has been successfully used to analyze various physical quantities, such as the spectral function~\cite{PhysRevResearch.4.043082}, mass distributions of the induced
fission~\cite{Huo2023}, and the electron-carbon scattering data~\cite{PhysRevC.110.025501}.

The dropout architecture with a single hidden layer can be recalled as follows,
\begin{eqnarray}\label{bnn1}
\hat{y}=\sigma(xz_{1}W_{1}+b)z_{2}W_{2},
\end{eqnarray}
where $\sigma$ depicts the nonlinear activation function, $x$ represents the input data, $W_{1}$ represents the weight matrix connecting the input layer to the hidden layer, $b$ denotes the bias vectors, and $W_{2}$ represents the weight matrix that links the hidden to output layer. $z_{1}$ and $z_{2}$ are binary vectors in the sampling process.
MC-dropout approach is driven by the Bernoulli distributed random variables in the multiplying hidden activations.
The random variables take the value of $1$ with probability parameter $p$ and $0$ for other cases~\cite{gal16,Gal2016}.
This means that the neuron is invalid if the corresponding binary variable takes $0$ in a given input.
In our calculations, the dropout probability parameter is chosen as 0.5.

\begin{figure}[htbp]
\centering
\includegraphics[scale=0.8]{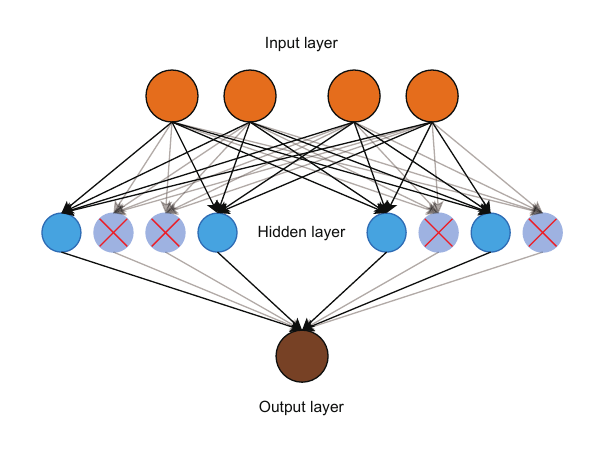}
\caption{Structure of the Monte Carlo dropout Bayesian neural network used in this work. The number of neurons in the input layer is $8$ or $10$. The $3$ hidden layers are employed, and the number of neurons in the corresponding hidden layer $1$ and $3$ are $28$ and $71$, respectively. The hidden layer $2$, namely the dropout layer, has $300$ neurons. This diagrammatic sketch just shows the dropout hidden layer, in which the different markers denote the random variables take value of $1$ with probability parameter $p$ and $0$ for other cases. The number of neurons in the output layer is $1$.} \label{fig1}
\end{figure}

The $L_{2}$ regularization weighted by some weight decay gives a minimization objective shown as follows~\cite{Huo2023,Wen2020},
\begin{eqnarray}\label{bnn3}
L_{\mathrm{dropout}}=\frac{1}{N}\sum_{n=1}^{N}||y_{n}-\hat{y}_{n}||_{2}^{2}+\lambda_{\mathrm{decay}}\sum_{i=1}^{L}(||W_{i}||_{2}^{2}+||b_{i}||_{2}^{2}).
\end{eqnarray}
The first term denotes the Euclidean loss function used in the neural network training and optimization process.
Where $\hat{y}_{n}$ represents the output of a neural network with $L$ layers and $y_{n}$ denotes the target value. $N$ is the number of training data sets. A regularization term, as shown in the second term, is often added to the optimization process.
Here, the weight matrices of the dimensions $K_{i}\times{K_{i-1}}$ are represented by $W_{i}$ and the bias vectors of the dimensions $K_{i}$ for each layer $i=1, 2, \cdots, L$ are indicated by $b_{i}$. The hyperparameter $\lambda_{\mathrm{decay}}$ in Eq.~(\ref{bnn3}) represents the regularization parameter~\cite{Huo2023}, which is used to reduce the risk of overfitting and its value is $3\times10^{-9}$. Meanwhile, the hyperparameters used in this work are identical for the constructed models (Fig.~\ref{fig1}). 

Shown in Refs.~\cite{Zhang10662945,Hüllermeier2021}, the predicted errors of the observables are mainly originating from epistemic uncertainty and aleatoric uncertainty in using machine learning approaches.
This means that the chosen input structure parameter sets should capture the underlying physical mechanisms as much as possible.
In this work, the experimental charge radii data of nuclei with $Z\geq20$ and $A\geq40$ are used to train the MC-dropout BNN~\cite{PhysRevLett.126.032502,ANGELI201369,LI2021101440}.
As mentioned in $Introduction$, the input structures for the MC-dropout BNN are consisted of the proton number ($Z$), mass number ($A$), isospin asymmetry degree ($I^{2}$), pairing effect ($\delta$), Casten factor ($P$), valence neutrons ($\nu_{n}$) and protons ($\nu_{p}$), quadrupole deformation ($\beta_{20}$), the hexadecapole deformation ($\beta_{40}$), and the profound shape-phase transition in $^{181,183,185}$Hg ($LI$). The ground-state properties about the quadrupole and hexadecapole deformation parameters are taken from Ref.~\cite{MOLLER20161}.

The final input structures for the MC-dropout BNN approach are $z=\{Z,$ $A$, $I^{2}$, $\delta$, $P$, $\nu_{n}$, $\nu_{p}$, $\beta_{20}$, $\beta_{40}$, $LI\}$, in which
\begin{eqnarray}\label{bnn4}
I^{2} &=&(\frac{N-Z}{A})^{2},\\
\delta &=& \frac{(-1)^{Z}+(-1)^{N}}{2},\\
P&=& \frac{\nu_{p}\nu_{n}}{\nu_{p}+\nu_{n}},\\
LI &=& \left\{
\begin{array}{c c}
1,&(Z,A)=(80,181),(80,183),(80,185)\\
0,&\mathrm{else}
\end{array}. \right.
\end{eqnarray}
As mentioned above, the traditional Casten factor $P$ is invalid along the isotopic chains of the proton magic numbers $Z=28$, $50$, and $82$ or the isotonic chains with the neutron numbers $N=28$, $50$, $82$, and $126$.
To avoid this puzzle, the modified Casten factor $P^{*}$ is proposed as follows,
\begin{eqnarray}\label{bnn5}
P^{*}&=& \ln\left(\frac{e^{\nu_{p}}e^{\nu_{n}}}{{\nu_{p}}+{\nu_{n}}}\right).
\end{eqnarray}
The input parameter sets that include the traditional Casten factor $P$ cannot describe the shell quenching phenomena well. In contrast to the traditional Casten factor $P$, the modified Casten factor $P^{*}$ can give more high accuracy in the training and validation sets, and the shell closure effect in the charge radii can be reproduced well. More details can be found in the following $Section$.
For the modified Casten factor $P^{*}$, the input structures can be redefined as $z=\{Z,$ $A$, $I^{2}$, $\delta$, $P^{*}$, $\nu_{n}$, $\nu_{p}$, $\beta_{20}$, $\beta_{40}$, $LI\}$.
Meanwhile, if the proton numbers and neutron numbers simultaneously occupy the fully filled shells, such as cases $(Z=28, N=28)$, $(Z=28, N=50)$, $(Z=50, N=50)$, $(Z=50, N=82)$, and $(Z=82, N=126)$, two kinds of Casten factor vanish completely.
In addition, the np formula~\cite{Nerlo-Pomorska:1994dhg} can also be chosen as our theoretical model to be refined by MC-dropout BNN:
\begin{eqnarray}\label{bnn6}
R_{\mathrm{np}}(Z,A)=r_{0}A^{1/3}\left(1-b\frac{N-Z}{A}+\frac{c}{A}\right),
\end{eqnarray}
where $r_{0}=0.966$ fm, $b=0.182$, and $c=1.652$~\cite{Bayram:2013jua}.

\section{Results and Discussion}
\label{sec3}
For convenience of discussion, we use $P^{*}$ to denote the results obtained by the MC-dropout BNN model with input neurons $z=\{Z,$ $A$, $I^{2}$, $\delta$, $P^{*}$, $\nu_{n}$, $\nu_{p}$, $\beta_{20}$, $\beta_{40}$, $LI\}$. We also give the results obtained by the MC-dropout BNN model with input neurons $z=\{Z,$ $A$, $I^{2}$, $\delta$, $P$, $\nu_{n}$, $\nu_{p}$, $\beta_{20}$, $\beta_{40}$, $LI\}$, which is denoted by $P$ for comparison.
Here, we should be mentioned that the constructed $P^{*}$ model is just used to avoid the distorted cases where the valence neutrons or protons are vanished.
In the training set, we use the $815$ experimental data given in Ref.~\cite{ANGELI201369}.
The more recent experimental data~\cite{PhysRevLett.126.032502,LI2021101440}, containing 118 data for nuclei with $Z\geq20$ and $A\geq40$, are used to test the predictive power of the constructed model in the validation set.
In contrast to Refs.~\cite{PhysRevC.105.014308,DONG2023137726}, the training data sets are reduced owing to the whole data of charge radii along nickel isotopic chain are transformed into the validation data set.

\begin{figure}[htbp!]
\centering
\includegraphics[scale=0.35]{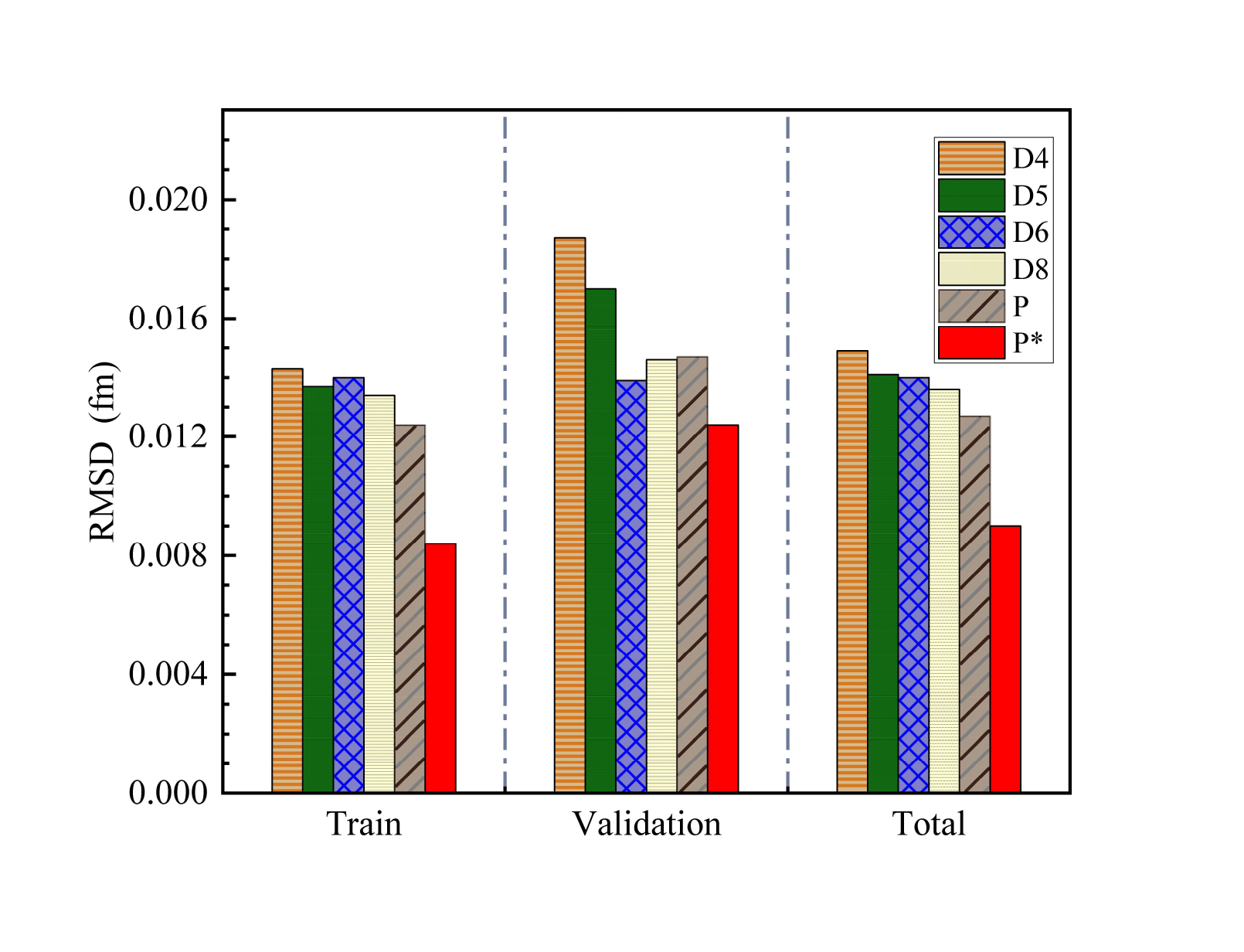}
\caption{Variations of the RMSDs for the train, validation, and total sets under the framework of D4~\cite{PhysRevC.105.014308}, D5~\cite{DONG2023137726}, D6~\cite{DONG2023137726}, D8, P, and P$^{*}$ models.} \label{fig0}
\end{figure}
To further quantify the predictive power, the root-mean-square deviation (RMSD) between the results obtained by the MC-dropout BNN model and the corresponding experimental data are calculated as follows,
\begin{eqnarray}\label{bnn7}
\sigma^{(\mathrm{T}, \mathrm{V}, \mathrm{TV})}=\sqrt{\frac{1}{N_{(\mathrm{T}, \mathrm{V}, \mathrm{TV})}}\sum_{i=1}^{N_{(\mathrm{T}, \mathrm{V}, \mathrm{TV})}}\left(R_{\mathrm{ch},i}^{\mathrm{theo}.}-R_{\mathrm{ch},i}^{\mathrm{exp}.}\right)^{2}}.
\end{eqnarray}
where $N_{\mathrm{T}}$, $N_{\mathrm{V}}$, and $N_{\mathrm{TV}}$ are the numbers of charge radii contained in the training (T), validation (V), and total (TV) data sets, and the subscript $i$ represents the $i$th nucleus in the given data sets.

To show the quantitative analysis, the influence of the input structure on the determination of the predictive power is discussed as follows. Meanwhile, the results obtained with the input structure $z=\{Z,$ $A$, $I^{2}$, $\delta$, $P^{*}$, $\nu_{n}$, $\nu_{p}$, $LI\}$ are also used to perform a comparison research, which is labeled as the D8 model. In contrast to the D8 model, the deformation parameters $\beta_{20}$ and $\beta_{40}$ are introduced into the input structures of the P model. As shown in Fig.~\ref{fig0}, the variations of the RMSDs for the train, validation, and total sets under the framework of the D4~\cite{PhysRevC.105.014308}, D5~\cite{DONG2023137726}, D6~\cite{DONG2023137726}, D8, P and P$^{*}$ models are plotted.
For the D4 model, the values of the RMSDs for the train, validation, and total sets are the largest with respect to other models. This can be understood easily, because the structure phenomena cannot be adequately captured in this model.
The D5 and D6 models give almost equivalent values of the RMSDs for the train and the total sets. In the validation set, the value of the RMSD obtained by the D6 model is less than that obtained by the D5 model due to the marked odd-even staggering of charge radii in the abrupt shape-phase staggering regions.

For the D4, D5, and D6 models, the Casten factor associated with the valence neutrons and protons is considered as input parameter.
However, shape deformation has not been taken into account in the validated protocol.
In recent study, the input sets $\{Z$, $A$, $\delta$, $\nu_{p}$, $\nu_{n}$, $\beta_{20}\}$ have been used in the training and validation data sets, the RMSDs have been reduced to $\sigma^{\mathrm{T}}=0.0106$ fm and $\sigma^{\mathrm{TV}}=0.0130$ fm~\cite{Zhang10662945}.
In contrast to the D6 model, the D8 model gives the lower RMSDs in the train and total data sets. For the validation data set, the RMSD value obtained by the D8 model is slightly larger than that obtained by the D6 model.
As demonstrated in Ref.~\cite{PhysRevC.88.011301}, the RMSD in charge radii can be reduced if the shape deformation is further incorporated into the calibrated protocol.
Here, it should be found that the P model further reduces the values of the RMSDs for the train and total data sets compared to the D8 model. Meanwhile, for the validation set, the value of the RMSD obtained by the D8 model is almost equivalent to the result obtained by the P model. 

\begin{table}[htb!]
\caption{Root-mean-square (rms) deviations of charge radii predicted by the $P$ and $P^{*}$ models.  }\label{tab1}
\doublerulesep 0.1pt \tabcolsep 12.8pt
\begin{tabular}{cccccc}
\hline
\hline
Model & $\sigma^{(\mathrm{T})}$ & $\sigma^{(\mathrm{V})}$ & $\sigma^{(\mathrm{TV})}$ \\
\hline
P        &0.0124 fm & 0.0147 fm & 0.0128 fm  \\
P$^{*}$  &0.0084 fm & 0.0124 fm & 0.0090 fm  \\
\hline
\hline
\end{tabular}
\end{table}
For the P$^{*}$ model, the values of the RMSDs for the train, validation, and total data sets are further reduced with respect to those obtained by the P model, or even the results obtained by the D4, D5, D6, and D8 models.
In Table~\ref{tab1}, the $\sigma^{\mathrm{T}}$, $\sigma^{\mathrm{V}}$, and $\sigma^{\mathrm{TV}}$ values of the P and P$^{*}$ models derived from the Eq.~(\ref{bnn7}) are shown, respectively.
In our calculations, the valence neutrons, valence protons, Casten factor, and the shape-phase staggerings in the limited region are simultaneously taken into account during the simulated process. Meanwhile, the shape deformation parameters $\beta_{20}$ and $\beta_{40}$ are also included.
The deduced value of $\sigma^{\mathrm{TV}}$ deriving from the $P$ model is equivalent to Ref.~\cite{Zhang10662945}, but less than the results shown in Refs.~\cite{PhysRevC.105.014308,DONG2023137726}.
The $\sigma^{\mathrm{TV}}$ value obtained by the $P^{*}$ model can be further reduced to be $0.0090$ fm.
Furthermore, the values of the RMSDs for the total set are sequentially reduced from the D4 model to the P$^{*}$ model. This means that more input quantities can reduce the epistemic uncertainty in the machine learning model. Consequently, the input structure has an influence on determining the predictive accuracy. Furthermore, it should be mentioned that the revised Casten factor is instructive to predicting the nuclear charge radii.

\begin{figure}[htbp!]
\centering
\includegraphics[scale=0.35]{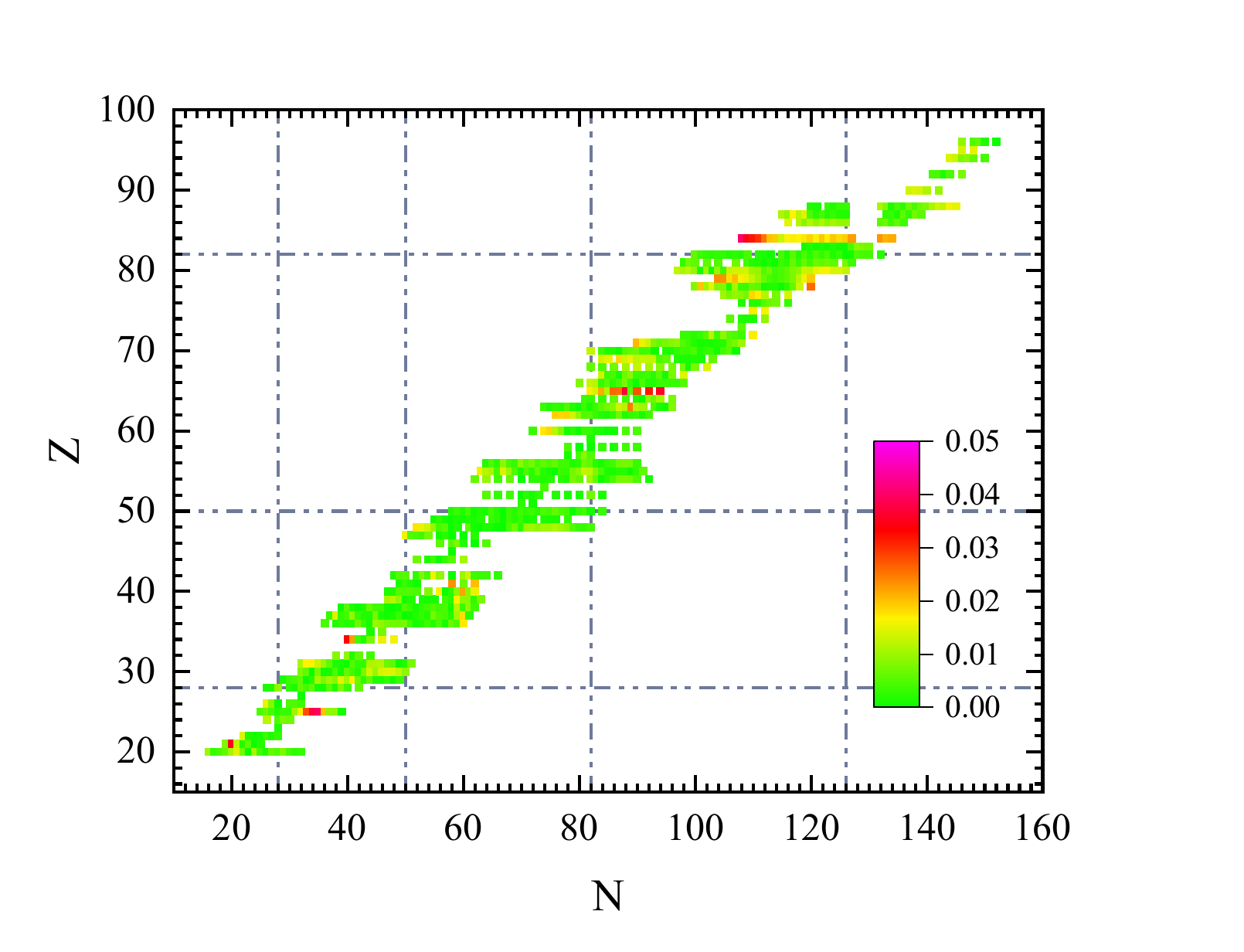}
\caption{Nuclear chart displaying the absolute differences (in units of fm) between the rms charge radii obtained by the $P^{*}$ model and the corresponding experimental data in the entire set. The results are obtained from the training and validation sets, respectively.} \label{fig2}
\end{figure}
Significant improvement has been achieved by the $P^{*}$ model in both the interpreted and the extrapolated predictions of nuclear charge radii compared to the $P$ model.
To visualize the details about the training and validation of the MC-dropout BNN approach, as shown in Fig.~\ref{fig2}, the difference between the calibrated rms charge radii with the $P^{*}$ model and the corresponding experimental data are depicted in the nuclear chart.
From this figure, it can be found that most of the experimental data can be reproduced well.
However, for some specific regions, the predicted values are heavily deviated from the experimental data.
As is well known, the inverse OES or the weakened OES can be evidently observed around Eu ($Z=63$)~\cite{Alkhazov1990} and At($Z=85$)~\cite{PhysRevC.99.054317} isotopes. In our calibrated protocol, the inverse odd-even staggering behavior has not been considered properly. This seems to lead larger deviations between the predicted rms charge radii and the corresponding experimental data. Meanwhile, larger deviations can also be encountered along manganese isotopes.
This means that more underlying mechanisms should be properly covered through the input structures.

\begin{figure*}[htbp!]
\centering
\includegraphics[scale=0.6]{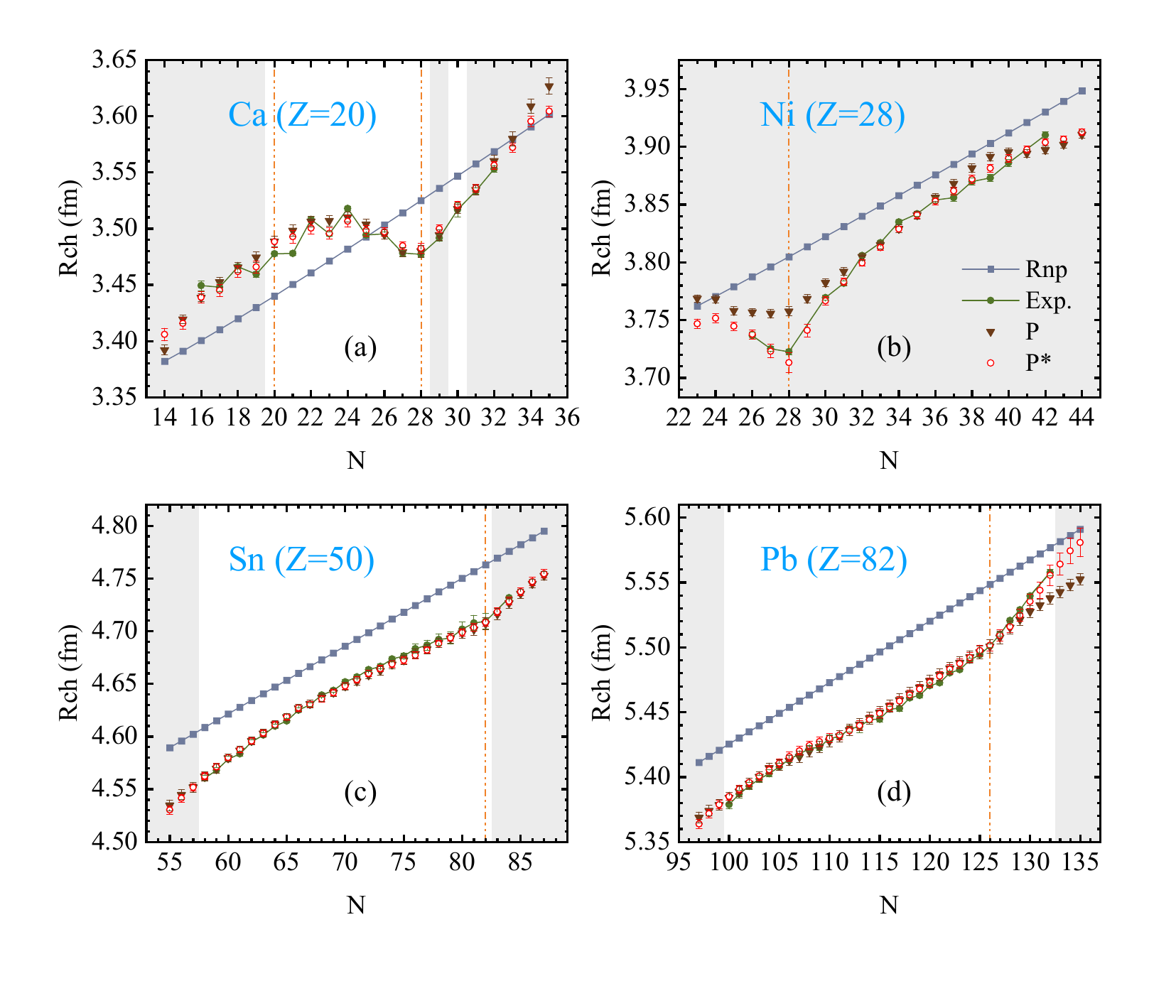}
\caption{Root-mean-square charge radii of calcium (a), nickel (b), tin (c), and lead (d) isotopes obtained by the $P^{*}$ and $P$ models compared to the experimental data~\cite{ANGELI201369,LI2021101440}. The results obtained by the np formula ($R_\mathrm{np}$) are also drawn for making a guide eye. The data covered by gray area are not contained in the training set, namely the pure predictions. The neutron magic numbers $N=20$, 28, 50, 82, and 126 are emphasized by the vertical lines. } \label{fig3}
\end{figure*}

Shell closure effects are naturally observed in the charge radii throughout the whole nuclear chart~\cite{ANGELI201369,LI2021101440,PhysRevLett.128.022502,PhysRevLett.129.132501,PhysRevLett.131.102501}.
This is attributed to the rather small isospin dependence of the spin-orbit interactions~\cite{PhysRevLett.74.3744}.
The relativistic mean field theory~\cite{Geng:2003pk,PhysRevC.102.024307,PhysRevC.109.064302} and the Fayans energy density functional model~\cite{PhysRevC.95.064328,PhysRevLett.122.192502} can reproduce the shell quenching phenomena in nuclear charge radii well.
However, as demonstrated in Ref.~\cite{PhysRevLett.128.022502}, the Skyrme density functional theory cannot describe the shell closure effects well.
This provides a potential signature to identify the predictive power of nuclear models in describing the local variations of the charge radii.
In Fig.~\ref{fig3}, the rms charge radii of calcium (a), nickel (b), tin (c), and lead (d) isotopes are depicted with both the $P^{*}$ and $P$ models, respectively.
As mentioned above, MC-dropout BNN model can provide the quantified uncertainty in the validated process.
Therefore the corresponding error bars are also shown in Fig.~\ref{fig3}.
For calcium isotopes, the $P^{*}$ and $P$ models give the similar trend of changes of charge radii. The slight deviations can be found at the neutron-deficient $^{34}$Ca and neutron-rich $^{54,55}$Ca isotopes. Meanwhile, charge radii of $^{39,43}$Ca isotopes can be reproduced well by the $P^{*}$ model compared to the $P$ model. Likewise, the systematical evolution of charge radii along tin isotopes can also be reproduced well by both the $P^{*}$ and $P$ models as shown in Fig.~\ref{fig3}~(c).

In Fig.~\ref{fig3}~(b), it is obviously found that the $N=28$ shell closure effect cannot be reproduced well by the $P$ model. This leads to heavily overestimated charge radii of the $^{54-56,58}$Ni isotopes. The same scenarios can also be found in the $^{65-68}$Ni isotopes, but with a slight deviation. Meanwhile, the charge radius of $^{70}$Ni obtained by the $P$ model is slightly underestimated compared to $P^{*}$ model.
Toward proton-rich regions, the pronounced deviations can be obviously exhibited between the predictive results obtained by the $P^{*}$ and $P$ models.
These larger deviations can also be encountered in the neutron-rich lead isotopes as shown in Fig.~\ref{fig3}~(d).
Here, it should be mentioned that the rapid increase in charge radii can be reproduced well by the $P^{*}$ model,
but the $P$ model fails to follow this trend beyond the neutron number $N=126$.
All of these results suggest that the $P^{*}$ model can provide a predictive power to describe the nuclear charge radii.
Therefore, an extrapolated simulation with high accuracy is necessary. In Table.~\ref{tab2}, charge radii of calcium, nickel, tin, and lead isotopes are shown explicitly. These results are encouraged to provide a theoretical guide for the experimental research.

The input parameters that include the modified Casten factor can describe the systematical evolution of nuclear charge radii well.
This can be easily found from the rms deviation of the training and validation data sets.
Furthermore, as shown in Fig.~\ref{fig3}, the shrunken phenomena in charge radii can be reproduced well by the $P^{*}$ model.
Actually, the Casten factor represents the neutron-proton correlation in describing the trend of changes of nuclear charge radii.
As demonstrated in Refs.~\cite{PhysRevC.53.1599,MILLER2019360,COSYN2021136526}, the neutron-proton correlation has an influence on determining the nuclear charge radii. This can also be found in Refs.~\cite{PhysRevC.102.024307,PhysRevC.109.064302} where the neutron-proton correlation derived from the difference between the neutron and proton Cooper pairs condensation can reproduce the shell quenching of nuclear charge radii as well.
All of these suggest that more appropriate interactions should be covered in describing the nuclear charge radii.
In particular, the difference of charge radii of mirror-pair nuclei can be used to ascertain the nuclear symmetry energy.
Thus, more available charge radii data are required in theoretical study.
\begin{table}[htb!]
\caption{The root-mean-square charge radii predicted by the $P^{*}$ and $P$ models, respectively. }\label{tab2}
\doublerulesep 0.25pt \tabcolsep 20.8pt
\begin{tabular}{ccc}
\hline
\hline
Nucleus  &   $P^{*}$ model  & $P$ model \\
\hline
$^{34}$Ca & 3.4059(56)  & 3.3920(44)   \\
$^{35}$Ca & 3.4156(53)  & 3.4193(37)   \\
$^{53}$Ca &  3.5721(40) & 3.5800(62)  \\
$^{54}$Ca &  3.5956(48) & 3.6089(65)  \\
$^{55}$Ca &  3.6046(47) & 3.6267(72)  \\
$^{51}$Ni &  3.7470(40) & 3.7687(33)  \\
$^{52}$Ni & 3.7517(40)  & 3.7682(35)  \\
$^{53}$Ni &  3.7448(35) & 3.7581(35)  \\
$^{57}$Ni & 3.7412(56)  & 3.7686(38)  \\
$^{69}$Ni & 3.8976(32)  & 3.8942(36)  \\
$^{71}$Ni & 3.9066(28)  & 3.9019(32)  \\
$^{72}$Ni & 3.9128(27)  & 3.9105(33)  \\
$^{105}$Sn & 4.5303(40)  & 4.5348(51)  \\
$^{106}$Sn & 4.5414(36)  & 4.5446(47)  \\
$^{107}$Sn & 4.5517(33)  & 4.5519(46)  \\
$^{133}$Sn & 4.7183(38)  & 4.7162(52)  \\
$^{135}$Sn & 4.7375(36)  & 4.7355(45)  \\
$^{136}$Sn & 4.7471(39)  & 4.7450(42)  \\
$^{137}$Sn & 4.7545(40)  & 4.7527(42)  \\
$^{179}$Pb & 5.3638(35)  & 5.3688(41)  \\
$^{180}$Pb & 5.3718(34)  & 5.3740(42)  \\
$^{181}$Pb & 5.3787(33)  & 5.3785(42)  \\
$^{213}$Pb & 5.5441(62)  & 5.5326(44)  \\
$^{215}$Pb & 5.5642(85)  & 5.5426(44)  \\
$^{216}$Pb & 5.5744(99)  & 5.5479(44)  \\
$^{217}$Pb & 5.5808(111)  &5.5525(44)   \\
\hline
\hline
\end{tabular}
\end{table}

\section{Summary and conclusions}
\label{sec4}
More input parameter sets physically motivated, such as proton number, mass number, isospin-asymmetry degree, pairing effect, shell closure effect, valence neutrons and protons, Casten factor $P$ or $P^{*}$, local shape-phase staggering phenomena, quadrupole deformation $\beta_{20}$, and higher order $\beta_{40}$ deformation, have been taken into account in the evaluated process.
Significant improvements can be achieved in training and validating data sets.
Especially, the $P^{*}$ model achieves superior performance, with rms deviations of $\sigma^{\mathrm{T}}=0.0084$ fm (train), $\sigma^{\mathrm{V}}=0.0124$ fm (validation) and $\sigma^{\mathrm{TV}}=0.0090$ fm (total).
The $P^{*}$ model can reproduce the shell quenching phenomena and the rapid increase of charge radii across neutron magic numbers.
This suggests that the $P^{*}$ model persists the predictive power when theoretical uncertainties are taken into account.
For nuclei located far from those contained in the training set, the $P^{*}$ model can still make reliable predictions.

This work presents the potential application of Bayesian neural network in exploring nuclear charge radii.
We should mention that more physically motivated features play an indispensable role in extrapolating the charge radii of nuclei far away from the $\beta$-stability line.
Although the Monte Carlo dropout Bayesian neural network provides rather small root-mean-square deviation in simulating the theoretical results, the larger deviation can also be encountered in some local regions.
This means that the underlying physical mechanism cannot be adequately captured in the training and validation data sets.
In addition, the number of training data sets has an influence on describing and predicting the nuclear charge radii.
Therefore, more underlying input structures should be taken into account properly in the proceeding investigation.

\section{Acknowledgements}
This work was supported by the Open Project of Guangxi Key Laboratory of Nuclear Physics and Nuclear Technology, No. NLK2023-05, the Central Government Guidance Funds for Local Scientific and Technological Development, China (No. Guike ZY22096024), the Natural Science Foundation of Ningxia Province, China (No. 2024AAC03015), and the Key Laboratory of Beam Technology of Ministry of Education, China (No. BEAM2024G04).





\end{document}